# Nanoscale Thermal Imaging of Dislocation-Mediated Heat Transport

Ruilin Mao[1], Bingyao Liu[1,4]✉, Jiaxin Liu[1,2], Xiaoyue Gao[1], Junping Luo[1], Fachen Liu[1,2], Ruochen Shi[1], Jiade Li[1], Jinlong Du[1]✉, Peng Gao[1,2,3,4,5]✉

1 International Center for Quantum Materials and Electron Microscopy Laboratory, School of Physics, Peking University, 100871 Beijing, P. R. China.

2 Academy for Advanced Interdisciplinary Studies, Peking University, 100871 Beijing, P. R. China.

3 Tsientang Institute for Advanced Study, Zhejiang, P. R. China.

4 Interdisciplinary Institute of Light-Element Quantum Materials and Research Center for Light-Element Advanced Materials, Peking University, 100871 Beijing, P. R. China.

5 Collaborative Innovation Center of Quantum Matter, 100871 Beijing, P. R. China.

✉e-mail: liubingyao@pku.edu.cn; jldu@pku.edu.cn; pgao@pku.edu.cn

## Abstract

Dislocations in crystalline materials are widely exploited to tailor the thermal conductivity of semiconductors and thermoelectrics, yet a critical gap persists: direct measurement of local thermal resistance at individual buried dislocations—along with its spatial extent—remains elusive due to the limitations of conventional thermal probes. Here, we use *in situ* scanning transmission electron microscopy–electron energy-loss spectroscopy to map nanoscale temperature distributions across a low-angle $SrTiO_3$ grain boundary with periodic dislocation arrays. Our results reveal a temperature drop of ~47 K across the dislocation array. The associated temperature-field distortions are concentrated near the dislocation cores, consistent with stronger local thermal resistance at these discrete sites rather than a uniformly distributed resistance along the array. We further identify a distinct two-scale heat transport characteristic near the dislocation array: core-dominated effects over ~4.8–6.2 nm and extended inter-core influences over ~10.3–14.3 nm. Atomic-scale structural and vibrational analyses further reveal core-associated atomic reconstruction and localized optical-phonon perturbations, providing a microscopic basis for the stronger local thermal resistance inferred near dislocation cores. These findings quantitatively resolve spatial heterogeneity of dislocation-mediated heat transport, uncover its atomic-scale mechanism, and provide a quantitative basis for defect-engineering—guiding the design of high-performance thermoelectrics, semiconductors, high-temperature structural alloys, and other functional crystalline materials.

## Main

Defect engineering has emerged as a powerful strategy for tailoring lattice thermal conductivity in semiconductors, thermoelectric materials and related functional materials[1–4]. As ubiquitous crystalline defects, dislocations play a pivotal role in governing transport properties in crystalline materials[5], especially thermal transport. For instance, dense dislocation arrays enhance bulk thermoelectrics performance[6,7], while dislocations in single-crystal group-III nitride films induce anisotropic thermal transport[8]. Yet, advancing from empirical defect tuning to predictive design requires quantifying the spatial scale of dislocation-induced thermal resistance and uncovering its microscopic origin[9,10]—a critical missing link in the field. Theoretically, classical models[11,12] focus primarily on long-range strain-field scattering with empirical cutoff radii, whereas atomistic studies highlight that core-localized vibrational anomalies[13] can also exert a strong influence. Despite these foundational theoretical frameworks, direct experimental evidence clarifying how dislocations modulate local heat flow and over what length scale this effect persists remains elusive.

Realizing such measurements, however, presents inherent technical challenges: conventional techniques such as time-domain thermoreflectance (TDTR) only provide microscale-averaged thermal conductivity[14] and lack nanoscale spatial resolution. Surface-sensitive methods, including scanning thermal microscopy[15] (SThM) and micro-Raman thermometry[16], are incapable of characterizing buried dislocations. Recent advances in scanning transmission electron microscopy-electron energy loss spectroscopy (STEM-EELS), however, now enable simultaneous mapping of structural distortions and vibrational responses at individual defects[17,18]. When combined with *in situ* heating within the microscope, these vibrational signatures further enable nanoscale mapping of local temperature fields[19,20]. This unique capability directly addresses the limitations of conventional probes, offering a pathway to resolve dislocation-mediated thermal resistance.

Here, we use a 5° low-angle grain boundary (LAGB) in $SrTiO_3$—featuring periodic dislocation arrays—as a model system, combining operando STEM-EELS with non-equilibrium molecular dynamics (NEMD) simulations. LAGBs provide an intrinsic length scale to distinguish whether thermal resistance is confined to dislocation cores or extends across inter-core regions. We find a total temperature drop of ~47 K across the dislocation array under heat flow, with resistance strongly localized at discrete cores (~4.8–6.2 nm) while inter-core regions remain transmissive (~10.3–14.3 nm). This two-scale behavior arises from core-localized structural reconstruction and reduced phonon density of states (PDOS) overlap. Our results resolve the spatial extent of dislocation-induced thermal resistance, validate core-dominated heat transport mechanisms, and lay the groundwork for predictive defect engineering of thermal conductivity.

We first investigated the thermal transport behavior through dislocations near a dislocation array using molecular dynamics (MD) simulations based on a machine-learned interatomic potential. The potential was trained on a density functional theory based dataset generated from ab initio molecular dynamics (AIMD) calculations, with an average force error of approximately 100 meV Å$^{-1}$ (**Fig. S1**). On this basis, we constructed a 5° LAGB model in $SrTiO_3$, whose dislocation core arrangement and local structure are in good correspondence with the experimental high-angle annular dark-field (HAADF) results, as shown in **Fig. 1a**.

In NEMD simulations, a temperature bias was applied across the two ends of the model to establish a steady-state heat flux, from which the temperature field near the dislocation array was obtained (**Fig. 1b**). The isotherms near the dislocation array are clearly non-planar and become markedly compressed at the dislocation cores, whereas they remain comparatively sparse in the regions between adjacent cores. This indicates that, under the same overall heat-flux driving force, the local temperature gradient is larger near the dislocation cores, corresponding to a stronger local thermal-resistance contribution. In other words, thermal resistance is not uniformly distributed along the dislocation array, but is spatially concentrated at discrete dislocation cores.

To further examine how the spatially heterogeneous distribution of thermal resistance reshapes local heat-flow pathways, we calculated the total heat-flux distribution near the dislocation array, as shown in **Fig. 1c**. The curl of the heat-flux field, $\nabla \times \mathbf{J}$, together with the corresponding streamlines and directional arrows, reveals pronounced positive and negative curl around the dislocation cores. This feature indicates that the heat flux is locally deflected and tends to bypass the cores rather than passing directly through them. This feature provides direct evidence that the dislocation cores induce nanoscale rerouting of heat flow. Furthermore, we calculated the spatial distribution of $\mathrm{d}J_x/\mathrm{dx}$ to resolve the variation in heat-flux components across the dislocation array, where $J_x$ is the heat-flux component along the average transport direction, as shown in **Fig. S2**. It decreases when approaching the dislocation core and rises thereafter. In contrast, within the inter-core regions, the $x$-direction heat flux increases before the dislocation array and decreases after it. This trend indicates local depletion and accumulation of the heat flux: heat is first diverted away from the dislocation cores and then redirected around them. Together, these results show that dislocation cores not only correspond to local thermal-resistance maxima, but also reorganize nearby heat-flow pathways on the nanoscale, whereas the inter-core regions behave more like comparatively transmissive channels across the boundary. Taken together, the temperature field and heat-flux distributions demonstrate pronounced spatial heterogeneity of thermal resistance: the dislocation cores act as stronger and more localized thermal bottlenecks, whereas the inter-core regions sustain more effective cross-boundary heat transport.

Using STEM-EELS combined with an *in situ* heating device, we experimentally investigated heat transport across the dislocation array on the 5° LAGB in $SrTiO_3$. The device geometry is shown in **Fig. 2a**. The left red region was coated with amorphous carbon and served as the heating electrode, with a typical resistance of 3 kΩ. Under an applied current of 5 mA, this amorphous carbon layer acted as the heat source on one side of the dislocation array. The right blue region was likewise coated with amorphous carbon and connected to the heat sink on the opposite side. Using this *in situ* device, we established a pronounced steady-state temperature gradient across the grain boundary.

The large field-of-view (FOV) HAADF image is shown in **Fig. 2b**, in which the dislocation array can be clearly identified. Local temperatures were extracted using the ln ($I_{loss}/I_{gain}$) intensity ratio within the 50–80 meV energy window, following the principle of detailed balancing. Representative data processing details and fitting errors are provided in **Fig. S3**. **Figure 2c** shows the temperature drop extracted across a 128 nm FOV together with linear fits to the two sides of the boundary. A clear temperature discontinuity is resolved at the dislocation array, yielding a total temperature drop of $\Delta T$=47 K, and the corresponding Kapitza length of 213 nm. The interfacial thermal conductance between the dislocation region and the bulk region was determined to be 56 MW $m^{-2}$ $K^{-1}$, based on the temperature drop at the interface, the temperature gradients on either side, and the intrinsic thermal conductivity of $SrTiO_3$; the detailed methodology is described in the **Methods** section.

To further resolve how this resistance is spatially distributed inside the dislocation array, we performed higher-magnification temperature mapping, as shown in **Fig. 2d**. The isotherms are clearly bent in the vicinity of the array rather than remaining planar, indicating substantial local redistribution of heat flow. In particular, the temperature contours become more compressed near dislocation cores, whereas the regions between neighboring cores exhibit a more diffuse temperature variation. This directly suggests that the thermal resistance is not uniformly distributed within the array, but instead varies on the scale of individual structural units.

To quantify this local heterogeneity, we extracted temperature profiles row by row along the y direction and fitted them using a Boltzmann-type sigmoidal function, as summarized in the stacking plot in **Fig. 2e**. From these fits, we obtained the thermal transition length and the local effective temperature gradient for each row (see Methods for details). **Figure 2f** shows that these two quantities exhibit an inverse relationship along the dislocation array. Dislocation core positions exhibit larger local temperature gradients and shorter transition lengths, while the regions between adjacent cores show smaller gradients and longer transition lengths.

More importantly, these two types of regions are associated with distinct characteristic length scales. At the dislocation-core positions, the thermal transition length is only ~4.8–6.2 nm, indicating a strong and highly localized thermal bottleneck.

By contrast, the inter-core regions display longer transition lengths of ~10.3–14.3 nm, corresponding to a weaker but more spatially extended modulation of heat transport. Taken together, these results show that thermal resistance in the dislocation array is strongly spatially heterogeneous: dislocation cores dominate the local resistance maxima, while the regions between neighboring cores remain more transmissive. The array as a whole produces a much larger temperature gradient than the crystalline regions on either side.

To further elucidate the origin of the enhanced local thermal resistance, we compared the local atomic structure and vibrational response in three representative regions: the dislocation-core region, the inter-core region between adjacent dislocation cores, and the bulk region far from the dislocation array. **Figure 3a** shows the experimental integrated differential phase contrast (iDPC) image near a dislocation core together with the corresponding atomic identification. Based on this atomic identification, we extracted the local in-plane bond geometry and obtained the Ti-O bond-length fitting result shown in **Fig. 3b**. Pronounced in-plane bond-length variations are concentrated near the dislocation core, directly indicating that the local coordination environment there deviates strongly from that in the surrounding lattice. **Figure 3c** further shows vibrational STEM-EELS spectra acquired from the core, inter-core and bulk region. For clarity, the main spectral features are labelled P2–P5, corresponding to representative optical phonon responses in $SrTiO_3$. Relative to the bulk, the inter-core region exhibits only modest spectral changes, whereas the core region shows a 10 meV blueshift of P4. These experimental observations reveal that the dislocation core exhibits both stronger local structural distortion and a distinct vibrational behavior.

To further understand these experimental signatures, we theoretically analyzed the local structure and vibrational states via atomistic modeling. As shown in **Fig. 3d**, the first-shell radial distribution function (RDF) peak in the core region shifts towards shorter bond lengths and becomes broader relative to the inter-core and bulk region. Consistently, the fitted in-plane bond lengths in **Fig. 3e** show a much stronger local distortion near the dislocation core than in the surrounding regions. The calculated in-plane projected PDOS in **Fig. 3f** reproduces the selective modification of the P4-related feature near the dislocation core, indicating that the local structural reconstruction induces vibrational modes distinct from the bulk phonon eigenstates. Combined with the thermal-vibration ellipsoid analysis in **Fig. S4**, these results support the presence of a localized vibrational mode near ~70 meV associated with the reconstructed structural units at the dislocation core.

Phonon features induced by atomic reconstruction are highly localized. **Figure 3g** shows that the P4 peak position exhibits a pronounced localized blueshift at the dislocation cores, confirming that the phonon anomaly is spatially confined rather than broadly distributed across the array. The spatial correlation between the P4 shift and $Ti^{3+}$ enrichment, together with the more extended $\varepsilon_{xx}$ strain field, further indicates that the phonon anomaly is more closely associated with local electronic/bonding reconstruction than with long-range strain alone (**Fig. S5**). **Figure 3h** further shows that the PDOS overlap decreases markedly within ~2 nm of the dislocation cores, dropping below 90% in the core region, whereas the inter-core region and bulk maintain much higher overlap. This indicates that the vibrational spectrum at the dislocation cores becomes significantly less compatible with the surrounding lattice, thereby hindering phonon transmission across regions. Taken together, these experimental and theoretical results show that local atomic reconstruction at the dislocation cores induces anomalous localized vibrational modes and reduces vibrational matching, thereby generating a stronger local thermal-resistance contribution, in agreement with the core-dominated local thermal bottlenecks observed in **Fig. 2d.**

This work establishes a spatially resolved picture of dislocation-mediated heat transport at the nanoscale. We estimate the thermal conductance across the dislocation array to be approximately 56 MW $m^{-2}$ $K^{-1}$ (Kapitza length ~213 nm), consistent with previous nanoscale SThM measurements of ceramic grain boundaries[21]. In contrast, a recent transient thermal grating study[22] reported substantially higher thermal conductivity, which may plausibly arise from its ~200 μm probe size: such macroscopic measurements tend to average bulk contributions and may mask intrinsic thermal resistance heterogeneity. Our nanoscale temperature mapping helps rationalize this discrepancy by showing that thermal resistance is not uniformly distributed along the dislocation array, but is concentrated near dislocation cores, with sharp core-dominated resistance over ~4.8–6.2 nm and a broader, weaker inter-core thermal modulation over ~10.3–14.3 nm.

This two-scale picture refines the classical Klemens–Carruthers framework for dislocation-mediated thermal transport[11,12], which treats dislocations mainly as long-range strain-field scattering sources, with cutoff-dependent regularization of the strain field and only simplified treatment of the core. While this framework may account for gradual inter-core thermal modulation, it cannot fully capture the sharp resistance peaks at individual cores, a distinction that may help rationalize prior discrepancies between bulk measurements and classical predictions in LiF[23] and InN[8]. These observations point to a refined picture in which dislocation-induced thermal resistance arises from complementary core-localized structural/vibrational reconstruction and extended strain-field scattering, with the former dominating the local thermal bottleneck.

Combined structural and vibrational characterization clarifies the origin of this core-localized thermal bottleneck. Ti–O bond distortions near the cores shift the P4 optical phonon feature of $SrTiO_3$ and reduce local–bulk PDOS overlap, weakening vibrational matching and concentrating thermal resistance at the cores. This mechanism is consistent with prior monochromated STEM-EELS studies of LAGBs and individual dislocations[17,18] and atomistic simulations of dislocation arrays[24-26]. More broadly, it aligns with atomic-scale evidence that dislocation cores can host emergent properties distinct from the bulk[27], reinforcing the role of core-specific physics in defect-mediated heat transport.

Beyond this mechanistic insight, our work establishes an experimental route to directly quantify the spatial extent of defect-induced thermal resistance. By combining *in situ* STEM-EELS thermometry with atomic-resolution structural characterization and vibrational spectroscopy, this approach enables local temperature, structure, and phonon response to be probed at the same buried defect under steady heat-flow conditions. Defining the spatial range of thermal resistance at individual defects provides a direct experimental benchmark for theories of dislocation-mediated heat transport and offers a physical basis for engineering heat flow through deliberate defect-structure design.

**Data availability**

The data that support the findings of this study are available from the corresponding author upon request.

**Code availability**

Custom scripts used for EELS temperature extraction, spectral fitting, and molecular-dynamics post-processing are available from the corresponding author upon reasonable request.

**Competing interests**

The authors declare no competing interests.

**Acknowledgements**

This work was supported by the National Natural Science Foundation of China (52125307 to Peng Gao and 12504198 to Jiade Li). Peng Gao acknowledges the support from the Xplorer Prize. We acknowledge the Electron Microscopy Laboratory of Peking University for the use of electron microscopes and the High Performance Computing Platform of Peking University for providing computational resources for the MD calculations.

**Author Contributions**

Peng Gao conceived the project. Ruilin Mao performed the STEM-EELS experiments, assisted by Fachen Liu and Jinlong Du, under the supervision of Peng Gao. Ruilin Mao and Bingyao Liu performed data processing and analysis, assisted by Ruochen Shi, Jiade Li, and Jinlong Du. Jiaxin Liu, Xiaoyue Gao, and Junping Luo assisted with iDPC experiments and atomic-position fitting. Ruilin Mao performed the molecular dynamics simulations and related theoretical analysis under the guidance of Peng Gao and Ruochen Shi. Ruilin Mao and Bingyao Liu wrote the manuscript under the supervision of Peng Gao. All authors discussed the results and contributed to the development of the manuscript.

## Methods

### Sample preparation and *in situ* heating

Commercial $SrTiO_3$ 5° low-angle grain-boundary specimens (Hefei Kejing Material Technology Co., Ltd.) were used to fabricate *in situ* heating devices. A TEM lamella containing the grain boundary was mounted onto a Si support chip (Fusion E-chips E-FEF01-A4, Protochips) with pre-patterned electrodes. Focused ion-beam milling was used to shape the specimen into a stripe geometry with a width of approximately 2 μm, and the grain-boundary region was thinned to below 50 nm. Amorphous carbon was deposited at one end of the stripe and connected to the electrode to serve as a local Joule heater, whereas the opposite end was kept thicker and connected to the support substrate to act as a heat sink, as shown in **Figure 2a**.

*In situ* heating was performed under constant-current conditions. Before EELS acquisition, the device was kept under the applied current for 1–2 h until the sample position and temperature distribution became stable. STEM-EELS mapping was then carried out under the steady thermal gradient.

### STEM-EELS acquisition

HAADF-STEM imaging was performed at 300 kV using an aberration-corrected FEI Titan Themis G2 microscope. Vibrational STEM-EELS measurements were conducted on a monochromated Nion U-HERMES200 microscope operated at 60 kV. The convergence and collection semi-angles were 35 mrad and 25 mrad, respectively, giving an energy resolution of 10–14 meV. For atomic-scale EELS mapping, the scan step was 0.125 nm and the dwell time was 1600 ms per pixel. The electron beam current used for vibrational EELS acquisition was kept below 0.1 nA to minimize beam-induced heating and irradiation damage.

### EELS data processing and temperature extraction

The vibrational EELS spectra were first aligned to correct energy drift. For heated datasets, the ZLP tail was removed using a vacuum-reference-assisted subtraction procedure. Specifically, a vacuum ZLP acquired under the same optical condition was aligned and affinely rescaled to minimize the least-squares residual between the measured spectrum and the reference elastic response before subtraction.

Local temperatures were obtained from the detailed-balance relation between the phonon-loss and phonon-gain signals,

$$\ln\left(\frac{I_{loss}(E)}{I_{gain}(E)}\right) = \frac{E}{k_B T}$$

where $I_{loss}(E)$ and $I_{gain}(E)$ are the energy-loss and energy-gain intensities at energy E, $k_B$ is the Boltzmann constant, and T is the local temperature. For each pixel, $\ln\left(\frac{I_{loss}(E)}{I_{gain}(E)}\right)$ was fitted as a linear function of energy within the selected vibrational energy window. The local temperature was calculated from the fitted slope $m$ as $T = 1/(k_B m)$.

For the core-loss data, the Ti-L edges were fitted by multiple linear least-squares (MLLS) method. The intrinsic Ti-L spectra of $LaTiO_3$ (for $Ti^{3+}$) and $SrTiO_3$ (for $Ti^{4+}$) form ref. 28 were used as reference spectra respectively.

**Interfacial thermal conductance**

The heat flux across the grain boundary was estimated from the temperature gradient in the bulk-like $SrTiO_3$ region,

$$J = -\kappa_{STO}\left(\frac{dT}{dx}\right)_{bulk}$$

where $\kappa_{STO}$ is the thermal conductivity of $SrTiO_3$ single crystal. Considering reported bulk single-crystal values of $SrTiO_3$ near room temperature, we used $\kappa_{STO}$=12 W $m^{-1}K^{-1}$ as an effective bulk thermal conductivity for estimating the interfacial thermal conductance[29]. The temperature drop across the grain boundary, $\Delta T_{GB}$ , was determined by linearly extrapolating the temperature profiles on both sides to the grain-boundary plane. The interfacial thermal conductance was then calculated as

$$G = \frac{J}{\Delta T}$$

**Transition-length analysis**

To quantify the spatial extent of the temperature transition near the grain boundary, temperature profiles were fitted using a Boltzmann-type sigmoidal function,

$$y = \frac{\Delta T}{1 + e^{(x-x_0)/\lambda}} + T_0$$

where $\Delta T$ is the fitted temperature change, $x_0$ is the transition center, $T_0$ is the low-temperature offset, and $\lambda$ characterizes the transition width. The transition length[30] was defined as $L_{tr}$=6$\lambda$, corresponding to the spatial range over which the fitted temperature changes from approximately 2.2% to 97.8% of $\Delta T$. The effective local temperature gradient was calculated as

$$\left|\frac{dT}{dx}\right|_{max} = \frac{\Delta T}{6\lambda}$$

This analysis was used to compare the thermal transition near dislocation cores and inter-core regions.

**Vibrational and structural analysis**

Characteristic vibrational peak positions were extracted by constrained multi-peak fitting of the processed EELS spectra. Atomic column positions in iDPC and HAADF-STEM images were refined by local two-dimensional Gaussian fitting. Sr and Ti columns were identified from HAADF contrast and used to analyze local lattice distortions and bond-length variations near the dislocation cores.

To quantify the vibrational similarity between different regions, the projected phonon density of states was compared using a normalized overlap,

$$O_i = \frac{\int g_i(\omega) g_{ref}(\omega) d\omega}{(\int g_i(\omega) d\omega)(\int g_{ref}(\omega) d\omega)}$$

where $g_i(\omega)$ is the local PDOS and $g_{ref}(\omega)$ is the reference PDOS. A larger $O_i$ indicates stronger vibrational similarity with the reference region.

**Atomistic simulations**

Molecular dynamics simulations were performed using GPUMD[31]. A neuroevolution potential for $SrTiO_3$ was trained using AIMD reference configurations sampled from bulk $SrTiO_3$ and related oxide phases over a broad temperature range. The potential was validated by comparing predicted energies, atomic forces and relaxed grain-boundary structures with the corresponding reference data and experimental STEM observations.

A 5° $SrTiO_3$ low-angle grain-boundary model was constructed by joining two oppositely rotated $SrTiO_3$ crystals, followed by energy minimization and finite-temperature relaxation. Equilibrium MD trajectories were used to calculate local projected phonon density of states from the velocity autocorrelation function.

For nonequilibrium MD simulations, Langevin thermostats were applied to hot and cold regions at the two ends of the model to establish a steady thermal gradient. The simulation cell was divided into spatial bins, and local temperatures were obtained from the atomic kinetic energy in each bin. The local heat-flux field was calculated using the virial–velocity correlation formalism implemented in GPUMD[32], in which the heat-current spectrum obtained from the Fourier-transformed virial–velocity correlation function was integrated over the full frequency range to yield the total heat current. The resulting temperature and frequency-integrated heat-flux fields were used to analyze the spatial distortion of heat flow near the dislocation array.

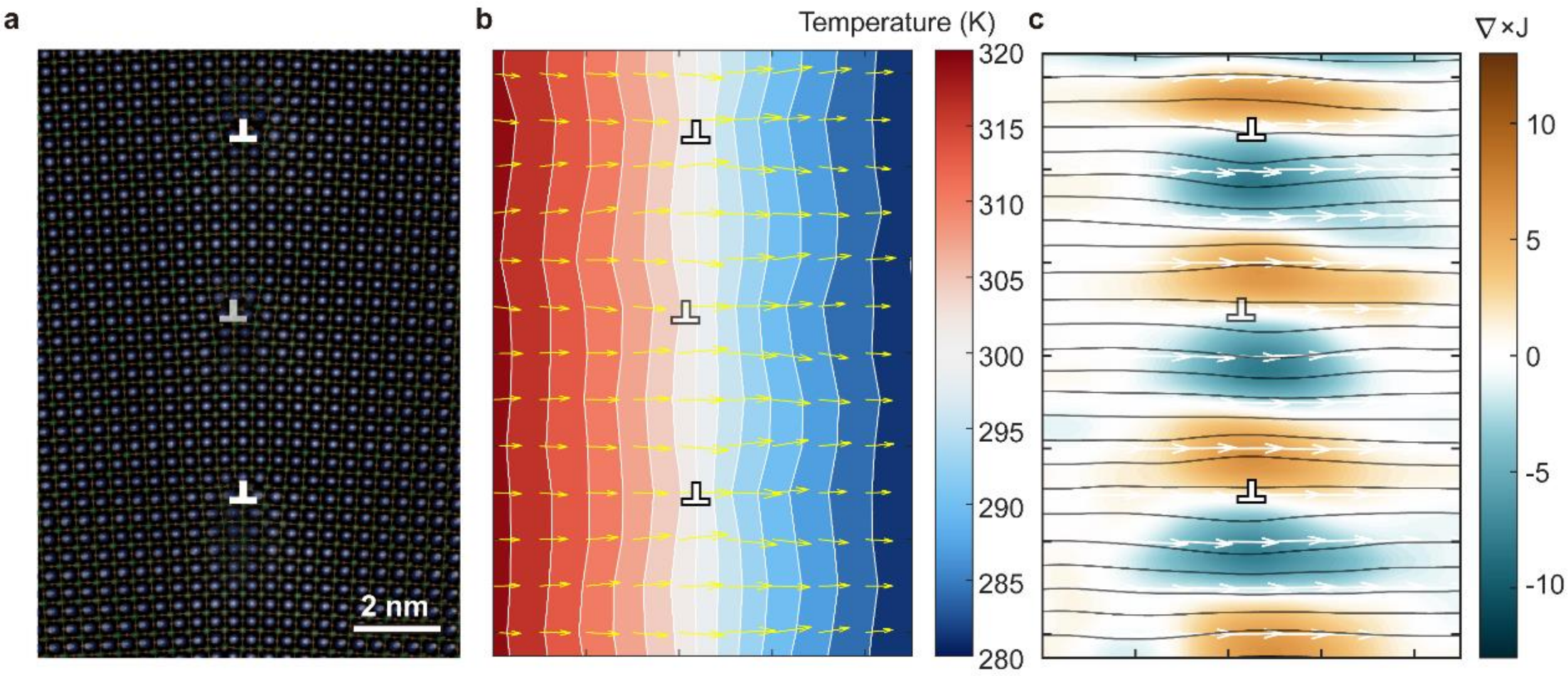

**Figure 1 | Simulated spatially heterogeneous heat transport across a dislocation array. a,** Overlay of the atomistic model on the experimental HAADF image of the dislocation array in a 5° low-angle grain boundary, showing the correspondence between the simulated structure and the periodically arranged dislocation cores observed experimentally. **b,** Temperature field obtained from NEMD showing a non-uniform thermal drop across the dislocation array. White contours denote isotherms. Yellow arrows denote the local direction and relative magnitude of the simulated heat flux. **c,** Curl of the heat-flux field, $\nabla\times J$, together with local streamlines and arrows.

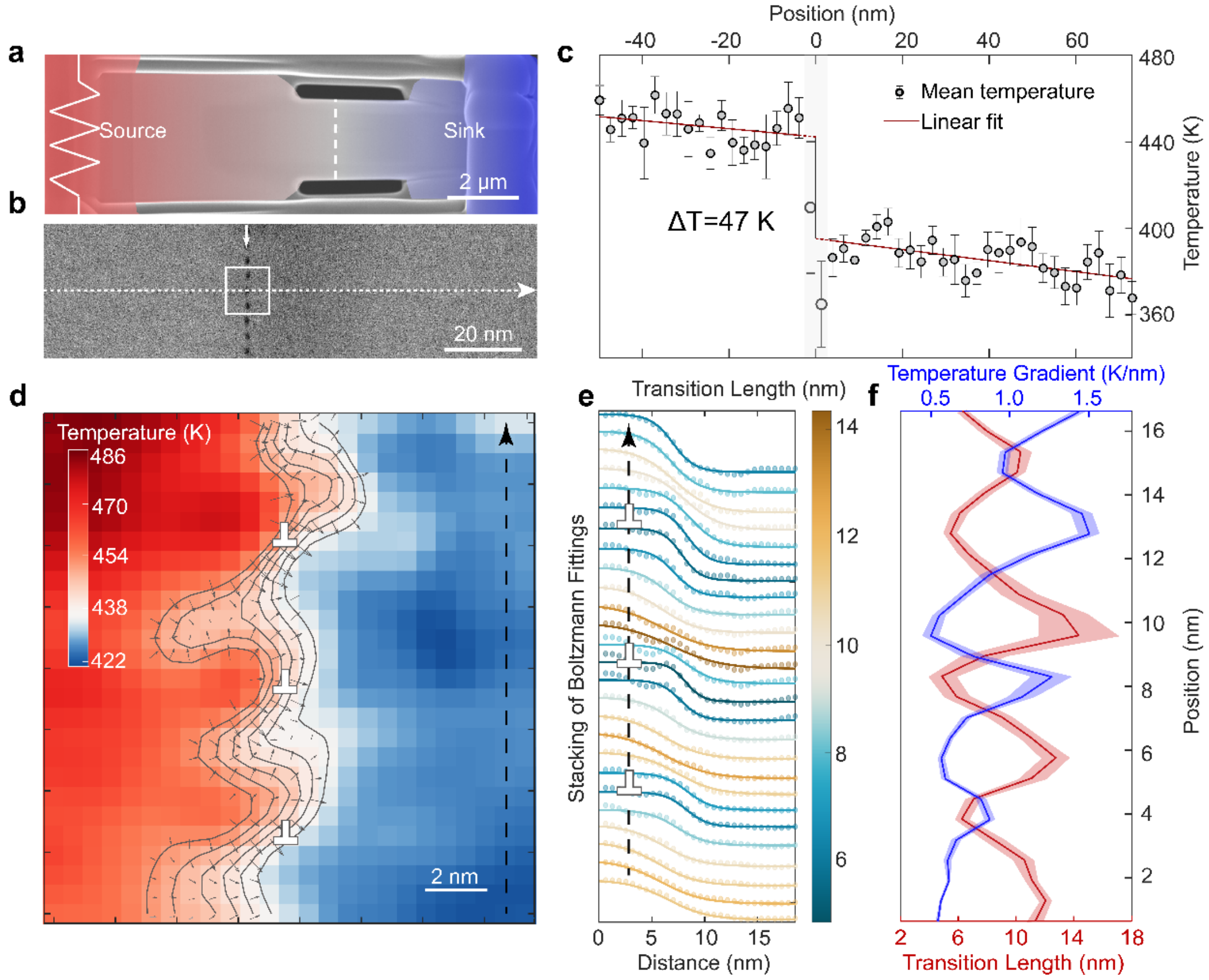

**Figure 2 | *In situ* STEM-EELS thermometry quantifies the effective thermal resistance and nanoscale heterogeneity of a dislocation array. a,** SEM image of the microfabricated *in situ* heating device used to establish a steady-state heat flux across the dislocation array. The amorphous carbon electrode on the left acts as the local heat source under applied current, while the opposite side is connected to the heat sink. **b,** Large FOV HAADF image of the measured region. **c,** Temperature profile measured across the dash arrow in (b). Linear fits to the temperature profiles on the two sides of the dislocation array reveal a temperature drop of ΔT=47 K across the dislocation array. **d,** Small FOV STEM-EELS temperature map near the dislocation array marked by the white box in (b). The overlaid contour lines highlight local bending and compression of the isotherms near the dislocation cores, indicating spatially heterogeneous thermal resistance. **e,** Row-by-row stacking of Boltzmann-type Sigmoidal fits to the local temperature profiles across the dislocation array. **f,** Spatial variation of the thermal transition length and effective temperature gradient along the dislocation array.

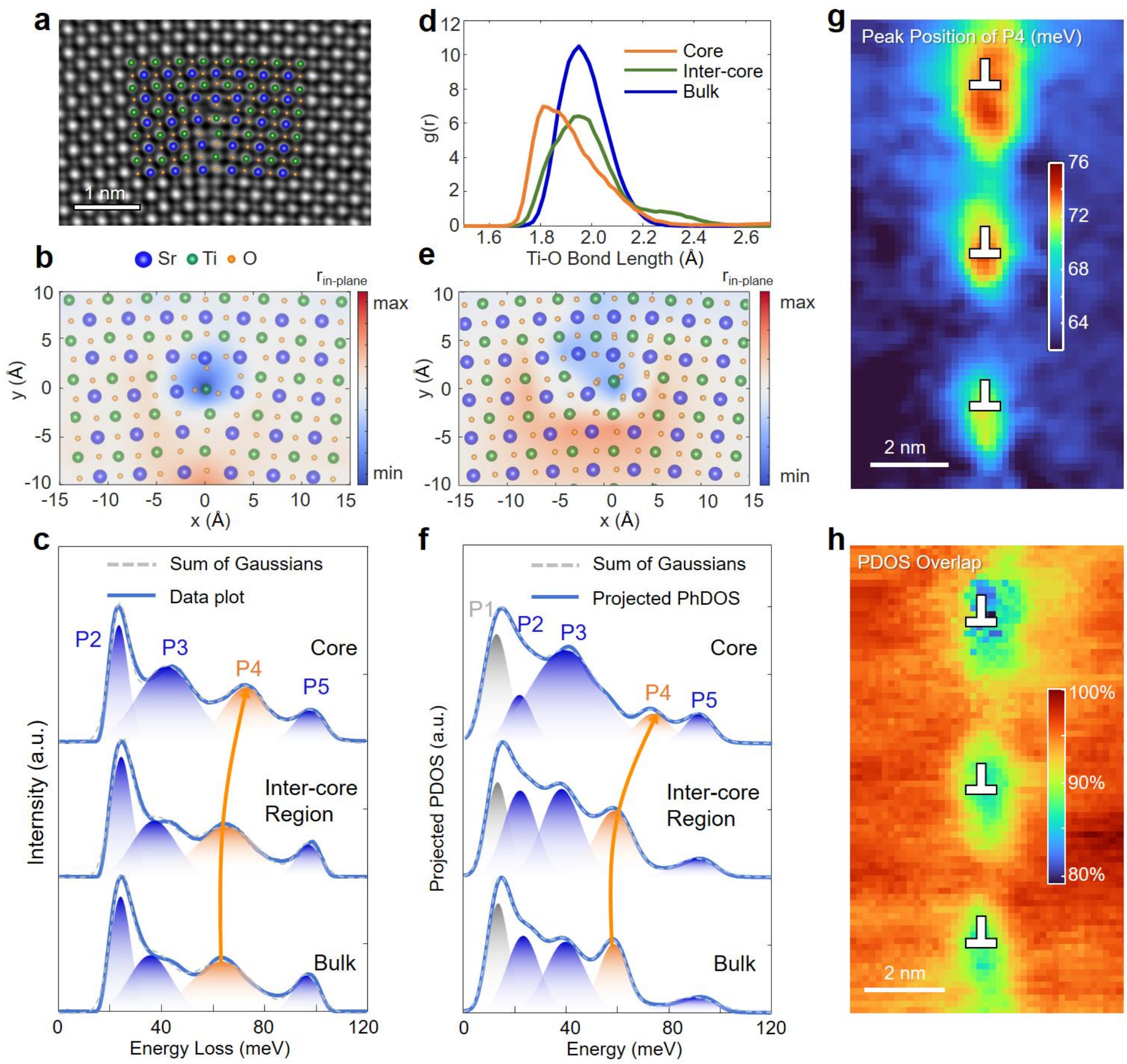

**Figure 3 | Local structural reconstruction and vibrational anomalies at dislocation cores correlate with the core-dominated thermal bottleneck. a,** iDPC image of the dislocation-array overlaid with the identified atomic species. **b,** Spatial map of the fitted in-plane Ti–O bond-length. **c,** Vibrational STEM-EELS spectra acquired from three representative positions, showing the most pronounced blueshift of P4 optical-phonon feature at the dislocation core. **d,** Simulated RDF, *g(r)*, of the first Ti–O coordination shell for the corresponding regions. **e,** Spatial map of the in-plane Ti–O bond-length derived from the relaxed atomistic model. **f,** Calculated local projected PDOS for the corresponding regions, showing the same P4 blueshift feature consistent with experiment. **g,** Experimental spatial map of the P4 peak position across the dislocation array. **h,** Experimental spatial map of the PDOS overlap across the dislocation array.

# Supplementary Information for Nanoscale Thermal Imaging of Dislocation-Mediated Heat Transport

Ruilin Mao[1], Bingyao Liu[1,4]✉, Jiaxin Liu[1,2], Xiaoyue Gao[1], Junping Luo[1], Fachen Liu[1,2], Ruochen Shi[1], Jiade Li[1], Jinlong Du[1]✉, Peng Gao[1,2,3,4,5]✉

1 International Center for Quantum Materials and Electron Microscopy Laboratory, School of Physics, Peking University, 100871 Beijing, P. R. China.
2 Academy for Advanced Interdisciplinary Studies, Peking University, 100871 Beijing, P. R. China.
3 Tsientang Institute for Advanced Study, Zhejiang, P. R. China.
4 Interdisciplinary Institute of Light-Element Quantum Materials and Research Center for Light-Element Advanced Materials, Peking University, 100871 Beijing, P. R. China.
5 Collaborative Innovation Center of Quantum Matter, 100871 Beijing, P. R. China.

✉e-mail: liubingyao@pku.edu.cn; jldu@pku.edu.cn; pgao@pku.edu.cn

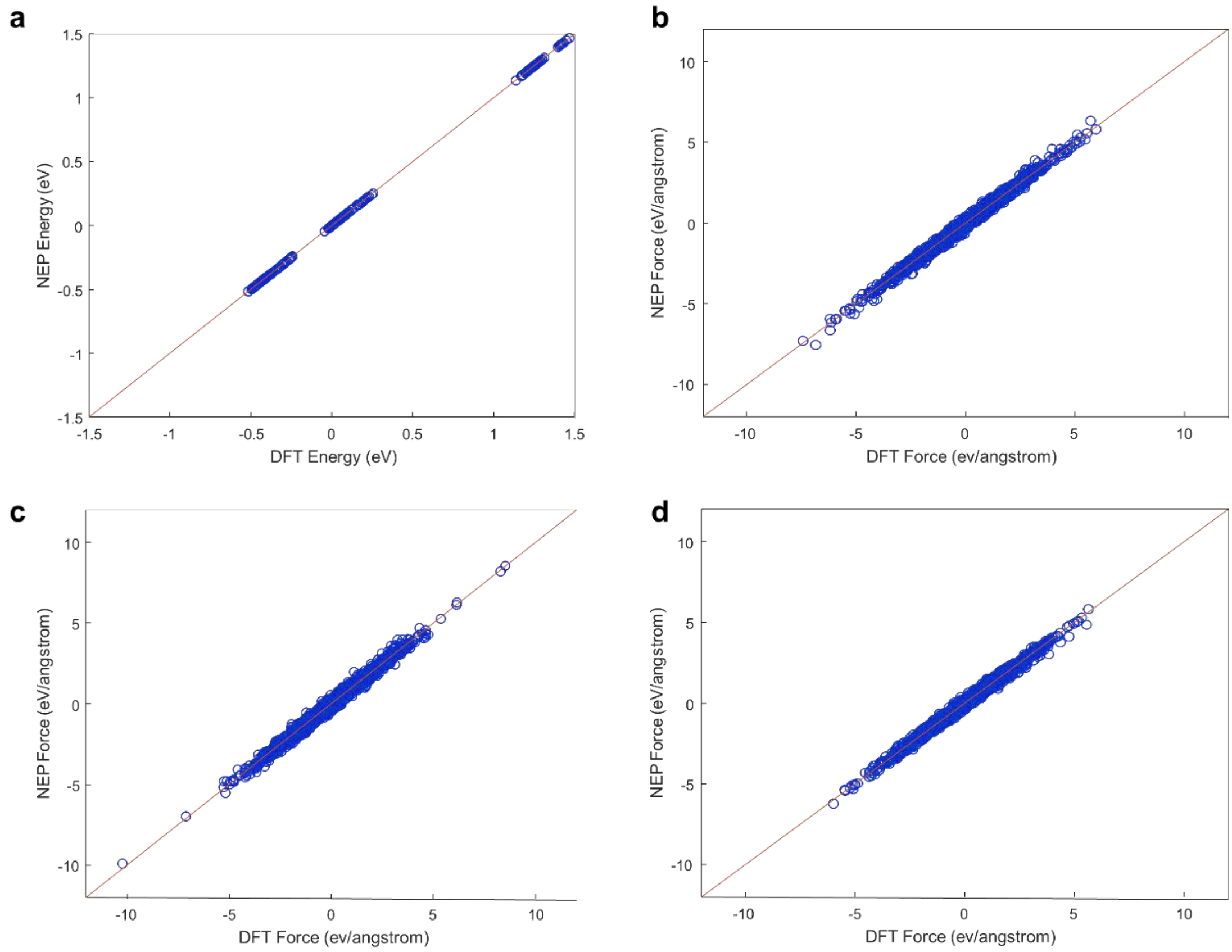

**Figure S1 | Validation of the NEP potential. a,** Parity plot of total energies predicted by the NEP potential versus DFT reference values. **b–d,** Parity plots of the x-, y- and z-components of atomic forces predicted by the NEP potential versus DFT reference values. The agreement confirms the accuracy of the potential for describing the local energetics and forces of the system.

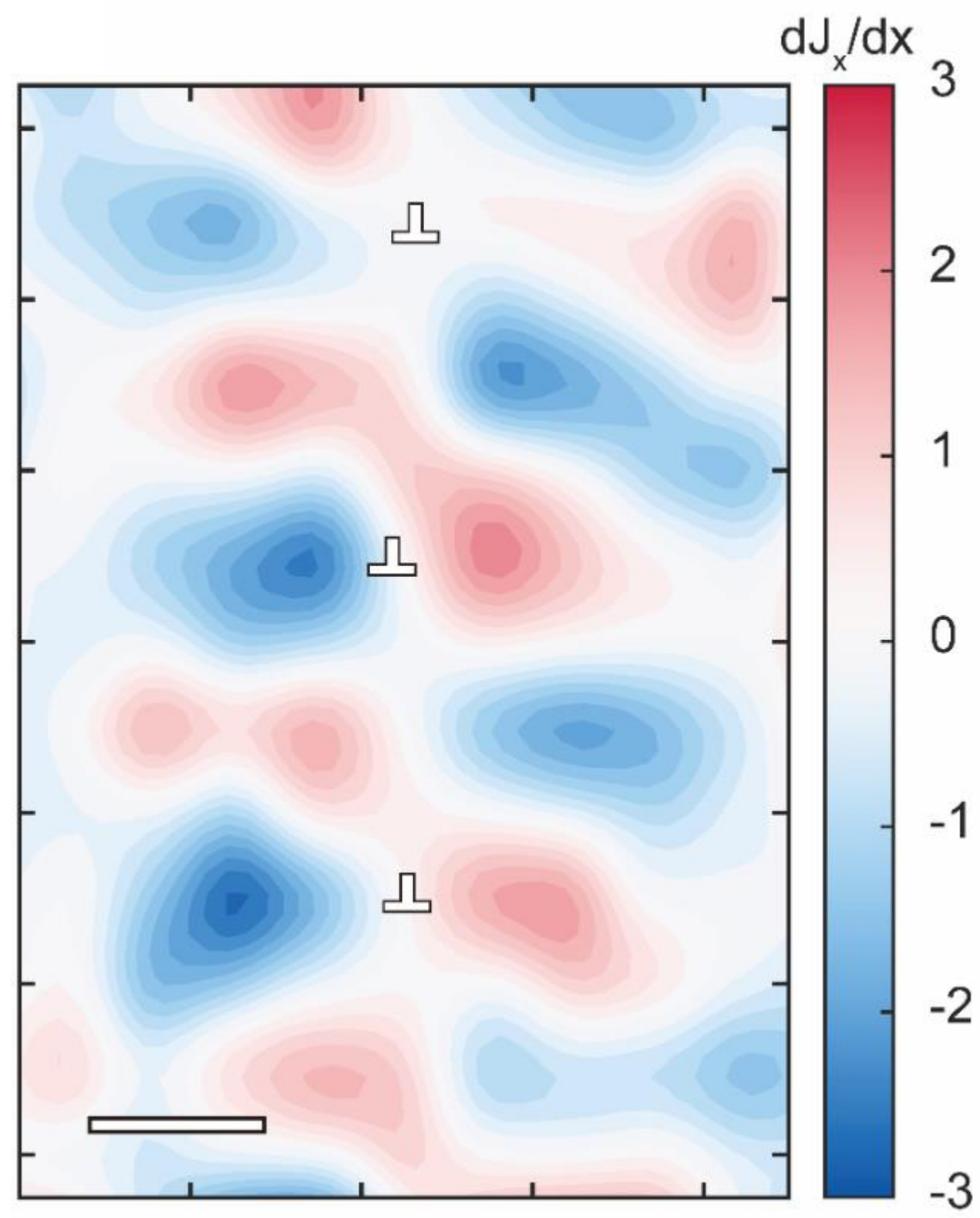

**Figure S2 | Spatial redistribution of the longitudinal heat-flux component near the dislocation array.** Spatial distribution of d$J_x$/dx, where $J_x$ is the heat-flux component along the average transport direction. Alternating positive and negative regions appear around the dislocation cores, indicating local accumulation and depletion of the longitudinal heat flux as heat is redistributed near the cores. This result further supports the nanoscale rerouting of heat flow around the dislocation cores and the comparatively smoother transport through the inter-core regions. Scale bar, 2 nm.

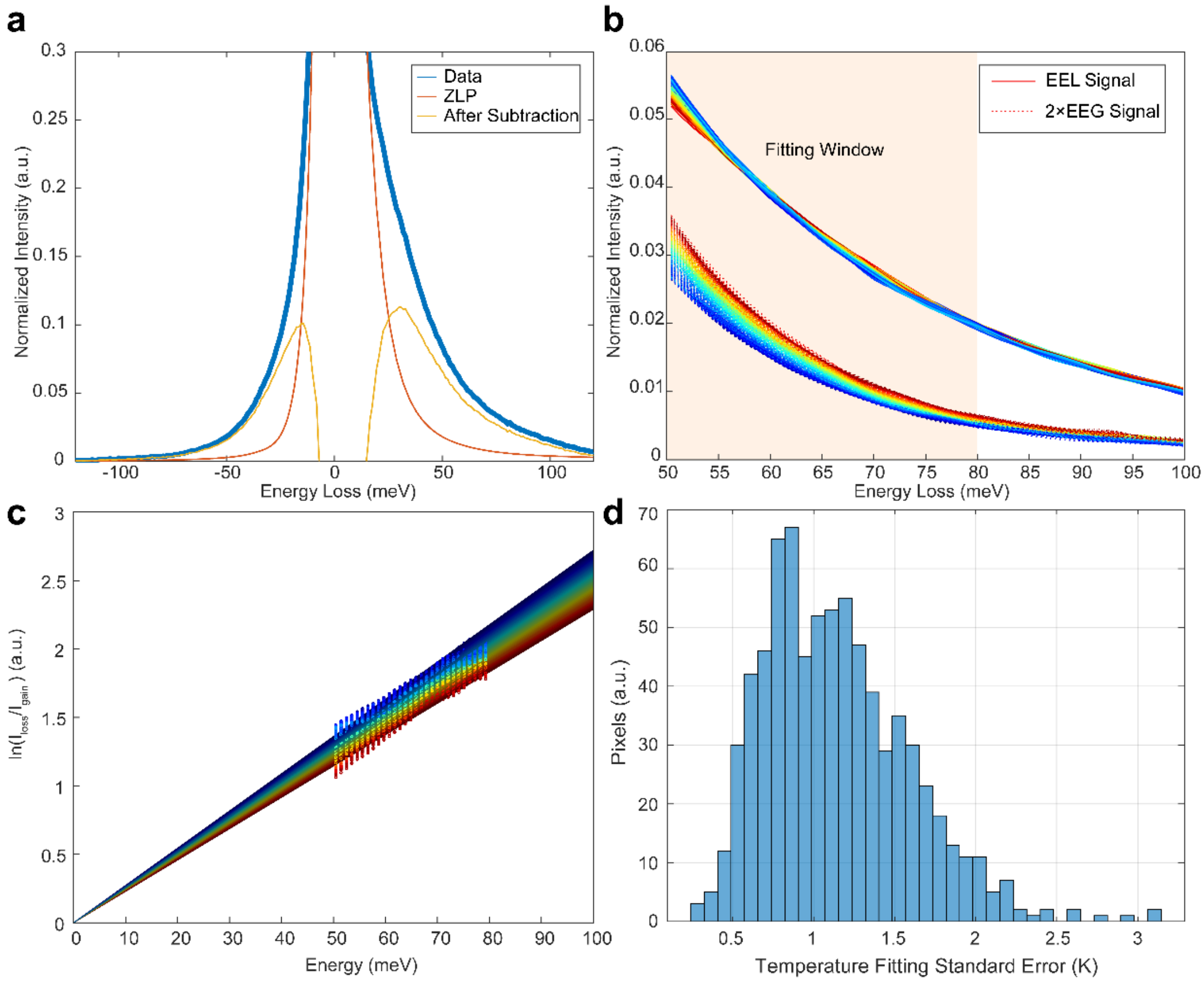

**Figure S3 | Temperature fitting based on the detailed-balance relation. a**, Representative spectrum showing the raw data, the fitted zero-loss peak (ZLP), and the separated energy-loss (EEL) and energy-gain (EEG) signals after ZLP-tail subtraction. **b**, EEL and EEG signals acquired at different temperatures, with the EEG signals scaled by a factor of 2 for clarity. For each temperature, the spectra were normalized by the integrated EEL intensity over the 50–80 meV energy-loss range. Colors indicate temperature from high (red) to low (blue). **c**, $\ln(I_{loss}/I_{gain})$ as a function of energy, with linear fits in the analysis window used to determine the local temperature. d, Histogram of the standard error from the linear fitting in c.

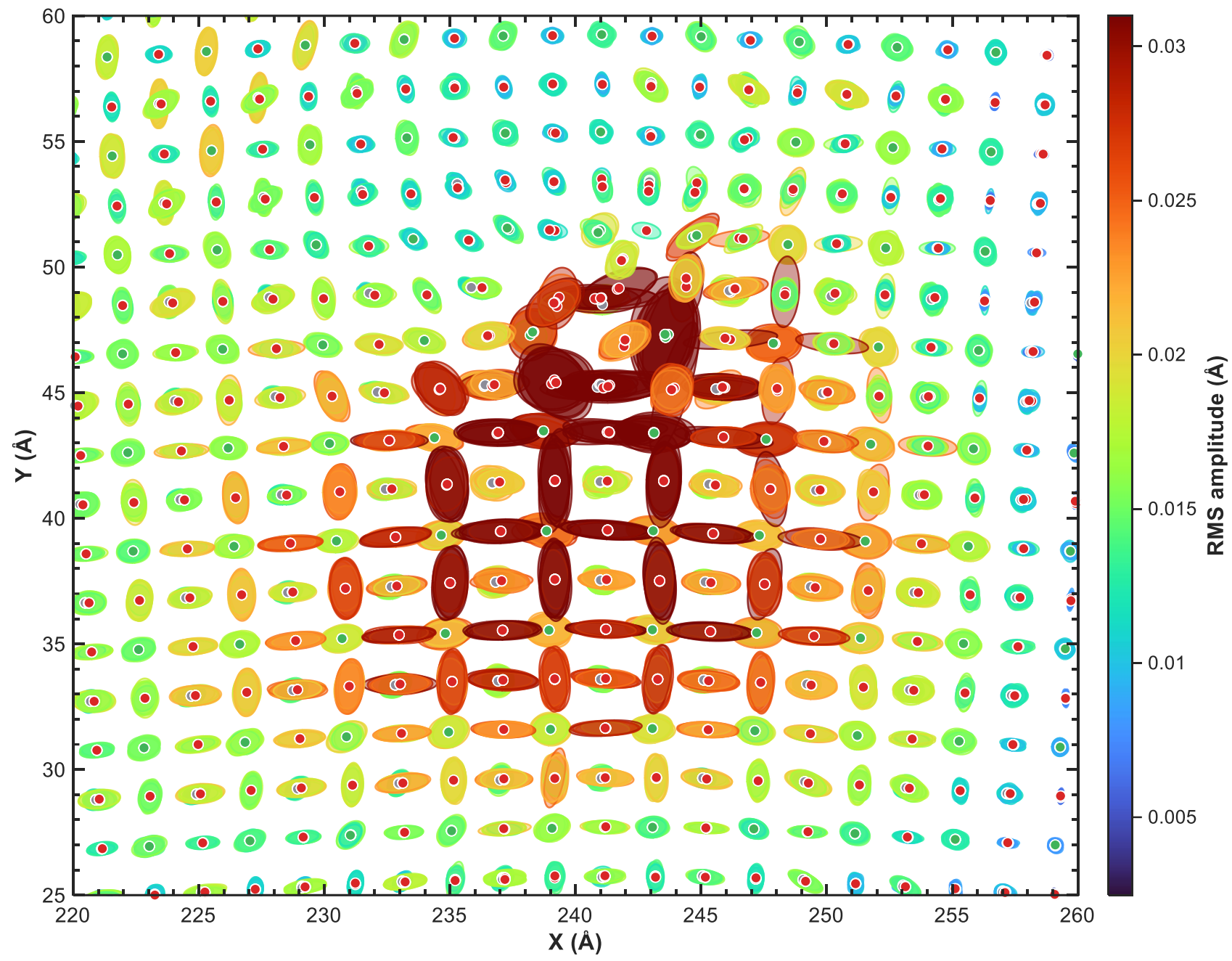

**Figure S4 | Thermal vibrational ellipsoids calculated for phonons in the 65–75 meV range.** Colors indicate the RMS vibrational amplitude. Enhanced and anisotropic vibrational amplitudes are localized near the dislocation core, supporting the presence of a core-associated localized vibrational mode.

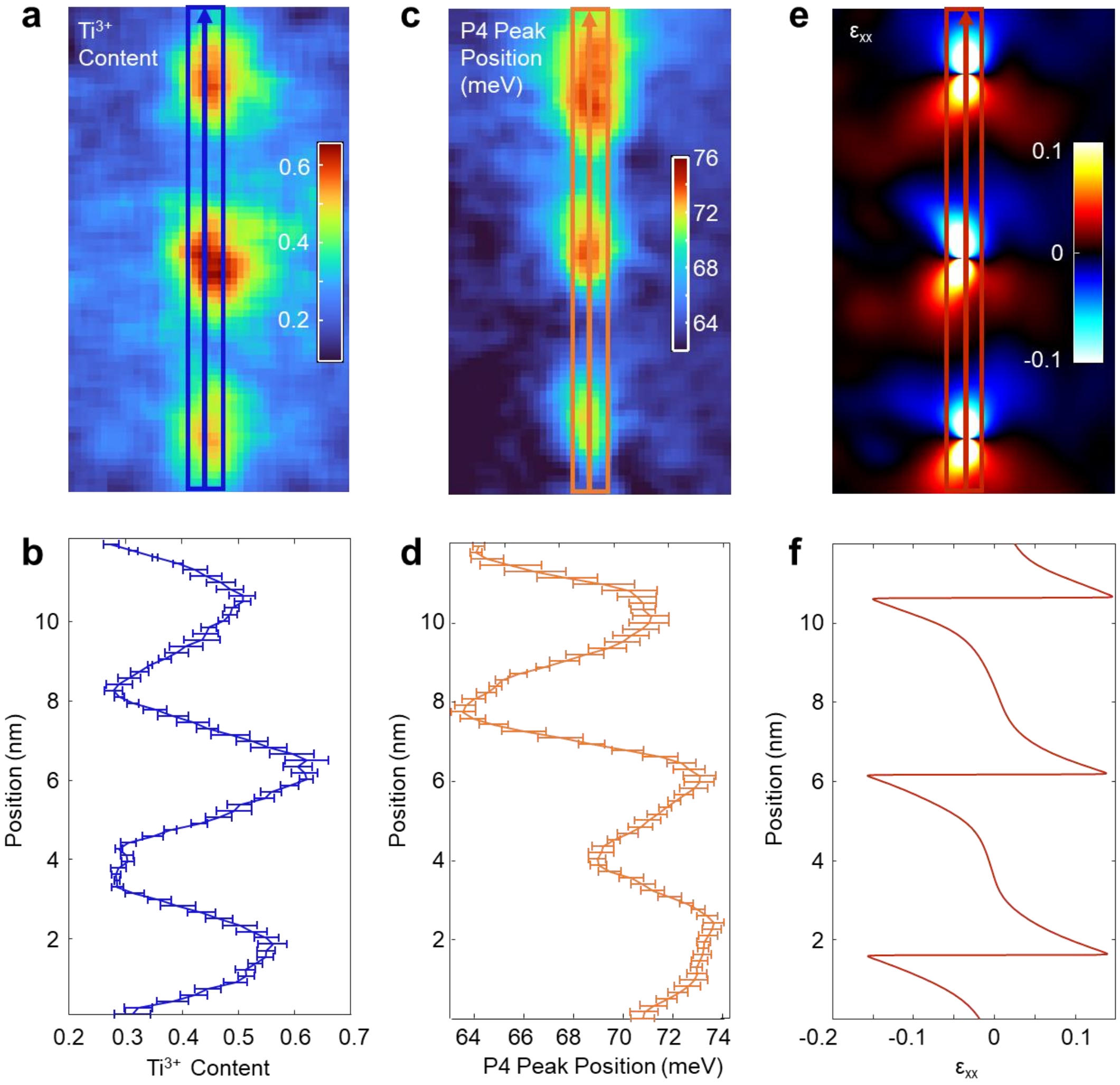

**Figure S5 | Spatial correlation between $Ti^{3+}$ content, phonon anomaly and strain near the dislocation cores. a,** Fitted map of the $Ti^{3+}$ spectral component extracted from Ti-$L_{2,3}$ core-loss EELS analysis, with representative core-loss spectra shown in **Fig. S6. b,** Corresponding line profile extracted from the boxed region in **a**. **c,** Fitted map of the P4 peak position. **d,** Corresponding line profile extracted from the boxed region in **c**. **e,** Local $\varepsilon_{xx}$ strain field obtained by geometric phase analysis. **f,** Corresponding line profile extracted from the boxed region in **e**. The P4 peak-position anomaly closely follows the $Ti^{3+}$ enrichment at the dislocation cores, whereas the strain field is spatially more extended.

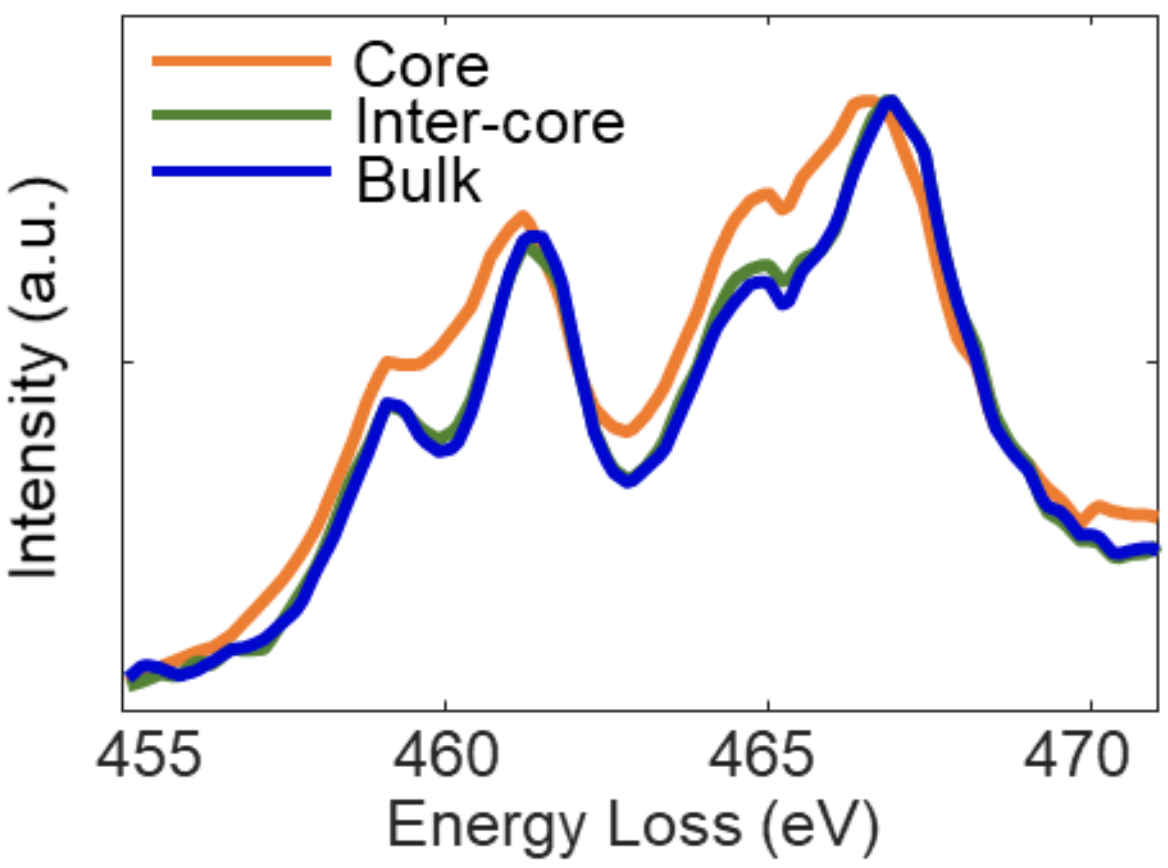

**Figure S6 | Ti-$L_{2,3}$ core-loss EELS spectra near the dislocation cores.** Representative Ti-$L_{2,3}$ core-loss EELS spectra acquired from the bulk-like region, inter-core region and dislocation-core region. Compared with the bulk-like region, the dislocation-core spectrum shows a reduced Ti-$L_{2,3}$ crystal-field splitting and modified near-edge fine structure, consistent with a local change in Ti–O coordination and an enhanced $Ti^{3+}$-like spectral contribution.